\documentclass[aps,prd,preprint,groupedaddress,nofootinbib]{revtex4}
\usepackage{graphicx}
\usepackage{dcolumn}

\begin{document}

\title{Model-Independent Bottom Baryon Mass Predictions in the $1/N_c$ Expansion}

\author{Elizabeth E. Jenkins}
\affiliation{Department of Physics, 9500 Gilman Drive,
University of California at San Diego, La Jolla, CA 92093}

\date{\today}

\begin{abstract}
Recent discoveries of the $\Xi_b$, $\Sigma_b$ and $\Sigma_b^*$ baryons at the Tevatron
are in good agreement with model-independent mass predictions made a decade ago based on a combined expansion in $1/N_c$, $1/m_Q$ and $SU(3)$ flavor symmetry breaking.  Using the new experimental data as input, mass predictions for the undiscovered bottom baryons  
$\Xi_b^\prime$, $\Xi_b^*$, $\Omega_b$ and $\Omega_b^*$ and for many unmeasured bottom baryon mass splittings are updated.  
The observed ground state charm baryons exhibit the mass hierarchy previously predicted by the $1/N_c$, $1/m_Q$ and $SU(3)$ flavor breaking expansion.
\end{abstract}

\maketitle

\def\clebsch#1#2#3#4#5#6{\left(\matrix{#1&#3\cr#2&#4\cr}\right.\left|
\matrix{#5\cr#6\cr}\right)}
\def\sixj#1#2#3#4#5#6{\left\{\matrix{#1&#2&#3\cr#4&#5&#6\cr}\right\}}
\def\xslash#1{{\rlap{$#1$}/}}
\def\Dsl{\hbox{/\kern-.6000em D}} 
\def\vabsq#1{\left|{\bf #1}\right|^2}
\def\dsl{\,\raise.15ex\hbox{/}\mkern-13.5mu D} 
\def\lqcd{\Lambda_{\rm QCD}}
\def\abs#1{\left| #1 \right|}
\def\vev#1{\left\langle #1 \right\rangle}
\def\bra#1{\left\langle #1 \right|}
\def\ket#1{\left| #1 \right\rangle}
\def\ltap{\ \raise.3ex\hbox{$<$\kern-.75em\lower1ex\hbox{$\sim$}}\ }
\def\gtap{\ \raise.3ex\hbox{$>$\kern-.75em\lower1ex\hbox{$\sim$}}\ }
\def\openone{\leavevmode\hbox{\small1\kern-4.2pt\normalsize1}}
\def\nn{\nonumber\\}
\def\Tr{{\rm Tr}}
\def\onebox{{\vbox{\hbox{$\sqr\thinspace$}}}}
\def\twobox{{\vbox{\hbox{$\sqr\sqr\thinspace$}}}}
\def\threebox{{\vbox{\hbox{$\sqr\sqr\sqr\thinspace$}}}}

The spectrum of baryons containing a single heavy quark was analyzed over a decade ago based on an expansion in $1/m_Q$, $1/N_c$, and $SU(3)$ flavor symmetry breaking~\cite{j1,j2}.  A hierarchy of mass splittings was
predicted for baryons containing a single heavy quark $Q$, and for differences of the heavy-quark spin-independent mass splittings of $Q=b$ baryons and $Q=c$ baryons.  Differences of  
spin-independent
mass splittings of heavy quark baryons and mass splittings of the octet and decuplet baryons also were
shown to exhibit an interesting hierarchy in the $1/N_c$ expansion.  From this analysis, all of the unknown charm and bottom baryon masses were predicted.  
Five of these baryons have since been discovered.  Table I gives the mass predictions of Ref.~\cite{j2} and the discovered masses; all of the observed masses are in good agreement with the theoretical predictions.  

In this paper, the recent precise measurements of the $\Xi_b$, $\Sigma_b$ and 
$\Sigma_b^*$ masses are used to update mass predictions for  
the remaining undiscovered bottom baryons $\Xi_b^\prime$, $\Xi_b^*$, $\Omega_b$ and $\Omega_b^*$ and other unmeasured bottom baryon mass splittings.  Since all charm baryons have been discovered, it also is possible to compare the entire theoretically predicted hierarchy of heavy quark baryon mass splittings given by the combined $1/N_c$, $1/m_Q$ and $SU(3)$ flavor breaking expansion with the experimental measurements for the case $Q=c$.  

The measured bottom baryon masses now are
\begin{eqnarray}
\Lambda_b &=& 5620.2 \pm 1.6 \ {\rm MeV}~\textrm{\cite{pdg}} ,\nn
\Xi_b &=& 5792.9 \pm 3.0 \ {\rm MeV}~\textrm{\cite{CDFxib}},\nn
\Sigma_b &=& 5811.5 \pm 1.7 \ {\rm MeV}~\textrm{\cite{CDFsigmab}},\nn
\Sigma_b^* &=& 5832.7 \pm 1.8 \ {\rm MeV}~\textrm{\cite{CDFsigmab}},
\end{eqnarray}
where the $\Xi_b$ mass is taken from the more precise CDF measurement, the quoted experimental values for $\Sigma_b$ and $\Sigma_b^*$ are the average mass of the two charged baryons, $\Sigma_b^{(*)} = {1 \over 2} \left( \Sigma_b^{+(*)} + \Sigma_b^{-(*)} \right)$, and statistical and systematic errors have been combined in quadrature. 
The experimental values of the charm baryon masses~\cite{pdg} are 
\begin{eqnarray}
\Lambda_c &=& 2286.46 \pm 0.14\ {\rm MeV}, \nn
\Sigma_c &=& 2453.56 \pm 0.16 \ {\rm MeV}, \nn
\Sigma_c^* &=& 2518.0 \pm 0.8\ {\rm MeV},  \nn
\Xi_c &=& 2469.5 \pm 0.3 \ {\rm MeV}, \nn
\Xi_c^\prime &=& 2576.9 \pm 2.1 \ {\rm MeV}, \nn
\Xi_c^* &=& 2646.4 \pm 0.9 \ {\rm MeV}, \nn
\Omega_c &=& 2697.5 \pm 2.6 \ {\rm MeV}, \nn
\Omega_c^* &=& 2768.3 \pm 3.0 \ {\rm MeV},
\end{eqnarray}
where the $\Omega_c^*$ mass is obtained from the measurement $(\Omega_c^* - \Omega_c) = 70.8 \pm 1.5 \ {\rm MeV}$~\cite{omegacstar} by BaBar.

Predictions for the four undiscovered bottom baryons, $\Xi_b^\prime$, $\Xi_b^*$ and $\Omega_b$ and $\Omega_b^*$, can be obtained using four mass relations.  The most accurate mass relations of Refs.~\cite{j1} and~\cite{j2} are used.  
The most suppressed mass combination in the $1/N_c$ expansion involving only baryons with a single bottom quark is
\begin{eqnarray}\label{mr1}
{1 \over 4} \left[ \left( \Sigma_b^* - \Sigma_b \right) - 2 \left( \Xi_b^* - \Xi_b^\prime \right) + \left( \Omega_b^* - \Omega_b \right) \right] &=& \pm 0.07\ {\rm MeV}
\end{eqnarray}
where $\pm 0.07$~MeV is the estimated size of the mass splitting based on theory~\cite{j1}.  This mass combination, which corresponds to the $J_Q^i \{ T^8, G^{i8} \}$ operator in the $1/N_c$ expansion, is
second order in $SU(3)$ flavor breaking, first order in $1/m_Q$ and suppressed by $1/N_c^2$.

A second very accurate mass relation
is obtained from the joint $1/N_c$ expansion for baryons containing a single bottom quark and baryons containing a single charm quark:
\begin{eqnarray}\label{mr2a}
&&{1 \over 6} \left[ \left( \Sigma_b + 2 \Sigma_b^* \right) - 2 \left( \Xi_b^\prime + 2 \Xi_b^* \right) + \left( \Omega_b + 2 \Omega_b^* \right) \right] \nn
&=& {1 \over 6} \left[ \left( \Sigma_c + 2 \Sigma_c^* \right) - 2 \left( \Xi_c^\prime + 2 \Xi_c^* \right) + \left( \Omega_c + 2 \Omega_c^* \right) \right]  \pm 0.3\ {\rm MeV}
\end{eqnarray}
where $\pm 0.3$~MeV is the theoretical estimate of the mass combination~\cite{j1}.  This mass combination corresponds to the $I_h^3\{T^8, T^8\}$ operator in the $1/N_c$ expansion, which is second order in $SU(3)$ breaking, suppressed by the heavy quark flavor symmetry-breaking parameter    
$(1/m_c - 1/m_b)\Lambda$, where $\Lambda$ is of order $\Lambda_{\rm QCD}$, and suppressed by $1/N_c^2$.  The charm baryon mass splitting in Eq.~(\ref{mr2a})
is equal to $-2.6 \pm 1.4$~MeV experimentally (see Table III).   The experimental error on the charm baryon mass splitting dominates the theoretical error of $\pm 0.3$~MeV of the mass relation Eq.~(\ref{mr2a}) at present.  A more accurate measurement of the charm baryon mass splitting would allow a
more accurate prediction of the analogous bottom baryon mass splitting based on Eq.~(\ref{mr2a}). 
An alternative mass relation for the bottom baryon mass combination on the left-hand side of Eq.~(\ref{mr2a}) is obtained from the joint $1/N_c$ expansion for baryons containing a single heavy quark and baryons containing no heavy quarks:
\begin{eqnarray}\label{mr2}
&&{1 \over 6} \left[ \left( \Sigma_b + 2 \Sigma_b^* \right) - 2 \left( \Xi_b^\prime + 2 \Xi_b^* \right) + \left( \Omega_b + 2 \Omega_b^* \right) \right] \nn
&=& {1 \over 3} \left[ {1 \over 4} \left( 2N - \Sigma - 3 \Lambda + 2 \Xi \right) + {1 \over 7} \left( 4 \Delta - 5 \Sigma^* - 2 \Xi^* + 3 \Omega \right) \right] \pm 1.5\ {\rm MeV}
\end{eqnarray}
where $\pm 1.5$~MeV is the estimated size of the mass combination based on theory~\cite{j1}.  This mass combination of heavy quark baryons and ordinary baryons corresponds to the $N_Q \{ T^8, T^8 \}$ operator in the $1/N_c$ expansion, which is second order in $SU(3)$ breaking and suppressed by $1/N_c^2$.   The combination of octet and decuplet masses on the right-hand side of the equation is equal to $-4.43$~MeV with negligible error.  Eq.~(\ref{mr2}) is a less accurate mass relation than Eq.~(\ref{mr2a}), but it currently yields a comparable extraction of the bottom baryon mass splitting given the experimental uncertainty of the charm baryon mass combination appearing in Eq.~(\ref{mr2a}).

A third very accurate mass relation is obtained from the joint $1/N_c$ expansion for baryons containing
a single bottom quark and baryons containing a single charm quark:
\begin{eqnarray}\label{mr3}
&&-{5 \over 8} \left( \Lambda_b - \Xi_b \right) + {1 \over {24}} \left[ 3 \left( \Sigma_b + 2 \Sigma_b^* \right) - \left( \Xi_b^\prime + 2 \Xi_b^* \right) - 2 \left( \Omega_b + 2 \Omega_b^* \right) \right] \nn &=& 
-{5 \over 8} \left( \Lambda_c - \Xi_c \right) + {1 \over {24}} \left[ 3 \left( \Sigma_c + 2 \Sigma_c^* \right) - \left( \Xi_c^\prime + 2 \Xi_c^* \right) - 2 \left( \Omega_c + 2 \Omega_c^* \right) \right] \pm 1.0\ {\rm MeV}
\end{eqnarray}
where $\pm 1.0$~MeV is the estimate of the mass combination based on theory~\cite{j1}.  
This mass combination corresponds to the $I_h^3 J_\ell^i G^{i8}$ operator in the $1/N_c$ expansion,
which is first order in $SU(3)$ breaking, suppressed by the heavy quark flavor symmetry breaking parameter $(1/m_c - 1/m_b)\Lambda$ and suppressed by $1/N_c^2$.
The charm baryon mass splitting is equal to $36.5 \pm 0.6$~MeV experimentally (see Table III).

The final mass relation is 
\begin{eqnarray}\label{mr4}
&&{1 \over 6} \left[ 3 \left( \Sigma_b^* - \Sigma_b \right) - \left( \Xi_b^* - \Xi_b^\prime \right) - 2 \left( \Omega_b^* - \Omega_b \right) \right] \nn &=&
{{Z_b m_c} \over {Z_c m_b}} \left({1 \over 6} \left[ 3 \left( \Sigma_c^* - \Sigma_c \right) - \left( \Xi_c^* - \Xi_c^\prime \right) - 2 \left( \Omega_c^* - \Omega_c \right) \right]\right),
\end{eqnarray}
which corresponds to renormalization improved $1/m_Q$ scaling of the corresponding charm baryon mass splitting.  The charm baryon mass splitting is measured to be $2.6 \pm 0.8$~MeV and the scale factor $({{Z_b m_c} / {Z_c m_b}})$ is equal to $0.24 \pm 0.05$~\cite{j2}, where $Z_b/Z_c = \left[ \alpha_s(m_b) / \alpha_s(m_c) \right]^{9/25}$.

Using the experimental input
\begin{eqnarray}
\left( \Sigma_b^* - \Sigma_b \right) &=& 21.2 \pm 2.5 \ {\rm MeV}, \nn
\left( \Lambda_b - \Xi_b \right) &=& -172.7 \pm 3.4 \ {\rm MeV}, \nn
{1 \over 3} \left( \Sigma_b + 2 \Sigma_b^* \right) &=& 5825.6 \pm 1.3 \ {\rm MeV},
\end{eqnarray} 
the four mass relations Eqs.~(\ref{mr1}), (\ref{mr2}), (\ref{mr3}), and~(\ref{mr4}) yield the predictions
\begin{eqnarray}\label{predictions}
\Xi_b^\prime &=& 5929.7 \pm 4.4 \ {\rm MeV},\nn
\Xi_b^* &=& 5950.3 \pm 4.2 \ {\rm MeV},\nn
\Omega_b &=& 6039.1 \pm 8.3 \ {\rm MeV}, \nn
\Omega_b^* &=& 6058.9 \pm 8.1\ {\rm MeV}.
\end{eqnarray}
(Use of Eq.~(\ref{mr2a}) rather than Eq.~(\ref{mr2}) yields comparable precision with slightly different central values.)

Specific linear combinations are predicted with greater precision:
\begin{eqnarray}\label{lincomb}
\left( \Xi_b^* - \Xi_b^\prime \right) &=& 20.6 \pm 1.9 \ {\rm MeV}, \nn
\left( \Omega_b^* - \Omega_b \right) &=& 19.8 \pm 3.1 \ {\rm MeV}, \nn
{1 \over 3} \left( \Xi_b^\prime + 2 \Xi_b^* \right) &=& 5943.4 \pm 4.2 \ {\rm MeV}, \nn
{1 \over 3} \left( \Omega_b + 2 \Omega_b^* \right) &=& 6052.3 \pm 8.0 \ {\rm MeV}.
\end{eqnarray}
The first two heavy quark spin-violating mass splittings are obtained using Eqs.~(\ref{mr1}) and~(\ref{mr4}), whereas 
the last two mass combinations are obtained using Eqs.~(\ref{mr2}) and~(\ref{mr3}). 
In addition, Eqs.~(\ref{mr1}) and~(\ref{mr2}) together imply
\begin{eqnarray}\label{mr9}
\left[ \frac 1 3 \left( \Sigma_b + 2 \Sigma_b^* \right) - \Lambda_b \right] - \left[ \frac 1 3 \left( \Xi_b^\prime + 2 \Xi_b^* \right) - \Xi_b \right] &=& 54.9 \pm 2.2 \ {\rm MeV}.
\end{eqnarray}
The first mass combination is square brackets is measured, see Table II, so Eq.~(\ref{mr9}) implies
\begin{equation}
\frac 1 3 \left( \Xi_b^\prime + 2 \Xi_b^* \right) - \Xi_b = 150.5 \pm 3.0 \ {\rm MeV} \ .
\end{equation}

Notice that
the extraction of the individual masses in Eq.~(\ref{predictions}) is dominated by the uncertainties of the
spin-averaged mass splittings in Eq.~(\ref{lincomb}).  These errors can be significantly reduced if the 
charm baryon mass splitting  ${1 \over 6} \left[ \left( \Sigma_c + 2 \Sigma_c^* \right) - 2 \left( \Xi_c^\prime + 2 \Xi_c^* \right) + \left( \Omega_c + 2 \Omega_c^* \right) \right]$ is measured more accurately, since mass relation Eq.~(\ref{mr2a}) then can be used to yield a much more accurate value of the analogous bottom baryon mass splitting.

It also is possible to extract the bottom baryon heavy quark spin-violating mass splittings by a simple rescaling of
the corresponding charm baryon mass splittings,
\begin{eqnarray}\label{mr5}
\left( \Sigma_b^* - \Sigma_b \right) &=& {{Z_b m_c} \over {Z_c m_b}} \left( \Sigma_c^* - \Sigma_c \right) ,\nn
\left( \Xi_b^* - \Xi_b^\prime \right) &=& {{Z_b m_c} \over {Z_c m_b}} \left( \Xi_c^* - \Xi_c^\prime \right), \nn
\left( \Omega_b^* - \Omega_b \right) &=& {{Z_b m_c} \over {Z_c m_b}} \left( \Omega_c^* - \Omega_c \right).
\end{eqnarray}
The experimental value $\left( \Sigma_b^* - \Sigma_b \right) = 21.2 \pm 2.5$~MeV and the theoretical extraction of $\left( \Xi_b^* - \Xi_b^\prime \right)$ in Eq.~(\ref{lincomb}) are more accurate than the values obtained using the rescaling Eq.~(\ref{mr5}).  The extraction $\left( \Omega_b^* - \Omega_b \right) = 17.0 \pm 3.6$~MeV following from Eq.~(\ref{mr5}) is only slightly less accurate than the value obtained in Eq.~(\ref{lincomb}).

The $1/N_c$ expansion for the baryon mass operator takes the form
\begin{equation}\label{opexpn}
M = \sum_{n=0}^{N_c} c^{(n)} {1 \over {{N_c}^{n-1}}}{{\cal O}^{(n)}},
\end{equation}
where the set of $n$-body quark operators ${{\cal O}^{(n)}}$ forms a complete and independent operator basis for the physical quantity considered~\cite{ncrefs}.  For baryons with $N_c$ valence quark lines, the set of independent operators ends with $N_c$-body operators.  
The operator bases for several different $1/N_c$ expansions involving heavy quark baryon masses were obtained in Ref.~\cite{j1}.  The matrix elements of all operators in the $1/N_c$ expansion are known, as is the leading $1/N_c$ dependence of each operator.  
Each operator in Eq.~(\ref{opexpn}), however, is multiplied by an unknown coefficient $c^{(n)}\left( {1 \over N_c} \right)$ which is order unity to leading order in the $1/N_c$ expansion.  Thus, the $1/N_c$ expansion predicts a scaling in $1/N_c$ of specific mass combinations corresponding to individual operators in the operator basis.  The $1/N_c$ expansion for heavy quark baryons generalizes to include an expansion in $1/m_Q$ and light-quark flavor symmetry breaking.  Each operator in the operator basis first occurs at a known order in heavy quark spin-flavor symmetry breaking and light-quark flavor symmetry breaking.  Thus, the combined $1/N_c$ expansion predicts a hierarchy of baryon mass splittings in the expansion parameters $1/N_c$, $(\Lambda_{\rm QCD}/m_Q)$ and $SU(3)$ flavor symmetry breaking.  The leading terms in the $1/N_c$ expansion are
\begin{equation}\label{leading}
M = N_c \Lambda \openone + N_Q m_Q + \cdots ,
\end{equation}
where, without loss of generality, the coefficients of the $\openone$ and $N_Q$ operators can be set equal to unity, thus defining $m_Q$ and $\Lambda$ for the expansion.     
The hierarchy of mass splittings of the various $1/N_c$ expansions studied in Ref.~\cite{j1} are summarized in Tables II-VIII.  Tables II-IV give the isospin flavor symmetry analysis of the mass hierarchy for baryons with fixed strangeness, whereas Tables V-VIII present the $SU(3)$ flavor symmetry-breaking analysis.  The mass relations Eqs.~(\ref{mr1})--(\ref{mr4}) used to predict the unmeasured bottom baryon masses in Eq.~(\ref{predictions}) are the most suppressed mass combinations in Tables V-VII. 

The measured heavy quark baryon mass splittings allow a detailed comparison of the theoretical hierarchy with the experimental data.
The experimental hierarchy is in excellent agreement with the $1/N_c$ hierarchy.
The $J_\ell^2$ mass splittings $\left[ \frac 1 3 \left( \Sigma_Q + 2 \Sigma_Q^* \right) - \Lambda_Q\right] $, $Q=c$ and $Q=b$,  in Table II are predicted to be equal to $2 \Lambda/ N_c$, independent of the heavy quark mass $m_Q$, up to a coefficient of order unity.
The experimental values of $210.0 \pm 0.5$~MeV and $205.4 \pm 2.1$~MeV correspond to coefficients very close to unity, for the value $\Lambda \sim 300$~MeV which follows from Eq.~(\ref{leading}).  Moreover, the difference of these $Q=c$ and $Q=b$ mass splittings (see Table III) is $4.6 \pm 2.2$~MeV, which is in excellent agreement with the theoretical prediction for
the $I_h^3 J_\ell^2$ mass splitting of $2 \Lambda/ N_c$ times the suppression factor $\left(1/2m_c - 1/2m_b \right) \Lambda /N_c \sim 1/45$ up to a coefficient of order unity.  The heavy-quark number dependent $N_Q J_\ell^2$ mass splitting
\begin{equation}
\left[ \frac 1 3 \left( \Sigma_Q + 2 \Sigma_Q^* \right) - \Lambda_Q\right]  - \frac 2 3 \left( \Delta - N \right),
\end{equation}
given in Table IV, is equal to $2 \Lambda / N_c$ times the suppression factor $1/N_c$ up to a coefficient of order unity.  The $Q=c$ and $Q=b$ mass splittings are more suppressed than this expectation, implying a coefficient significantly less than unity prediction of the $1/N_c$ expansion in this instance.  The heavy quark spin-dependent $J_Q \cdot J_\ell$ mass splittings $\left( \Sigma_Q^* - \Sigma_Q \right)$, $Q=c$ and $Q=b$, in Table II also are in excellent agreement with the theoretical prediction of $3 \Lambda /m_Q$ times a coefficient of order unity.  The $1/m_Q$ scaling of the mass splittings is evident.  There is further evidence for the $1/N_c$ hierarchy in the $SU(3)$ flavor symmetry-breaking analysis of Table V.  The $T^8$ mass splittings
$(\Xi_Q - \Lambda_Q)$, $Q=c$ and $Q=b$, compare favorably with the theoretical prediction $\frac {\sqrt{3}} 2 \left(\epsilon \Lambda_{\chi} \right)$, where $\Lambda_\chi \sim 1$~GeV and $\epsilon \sim 0.2 - 0.3$, times a coefficient of order unity.  Notice that the $J_\ell^i G^{i8}$ mass splitting is predicted to be suppressed relative
to the $T^8$ mass splitting by the factor $5/ 8 N_c$; the experimental $Q=c$ splittings
in Table V of $36.5 \pm 0.6$~MeV and $183.0 \pm 0.3$~MeV, respectively, show a relative suppression factor in excellent agreement with this expectation.  The flavor-$\bf 27$ mass splittings $\{ T^8, T^8 \}$ and $J_Q^i \{ T^8, G^{i8} \}$, suppressed by two powers of $SU(3)$ flavor
symmetry breaking, are measured for $Q=c$ baryons, but with significant error bars; a better experimental determination is necessary for a meaningful comparison with the theoretical hierarchy.  Most of the $SU(3)$ flavor-breaking mass splittings in Table V for $Q=b$ baryons are not yet determined experimentally, so the hierarchy of differences of charm and bottom baryon heavy-quark spin-independent mass splittings in Table VI is largely untested.  The $I_h^3 T^8$ mass splitting,
\begin{equation}
\left( \Xi_c - \Lambda_c \right) - \left( \Xi_b - \Lambda_b \right),
\end{equation}
is measured, and is consistent with the theoretical prediction of a suppression factor of
$\left(1/2m_c - 1/2m_b \right) \Lambda /N_c \sim 1/45$ relative to the individual $T^8$ mass splittings $\left( \Xi_Q - \Lambda_Q\right)$, $Q=c$ and $Q=b$, given in Table V.
The theoretical prediction is $\sim 5$~MeV, whereas experimentally the mass splitting is $10.3 \pm 3.4$~MeV.  The $N_Q T^8$ mass splittings of Table VII are predicted to be suppressed by a factor of $1/N_c$ relative to the $T^8$ mass splittings $\left( \Xi_Q - \Lambda_Q \right)$, $Q=c$ and $Q=b$, of Table V.  A suppression of this level is evident.  Finally, Table VIII lists the percentage accuracy of the mass splittings in the 
$SU(3)$ flavor symmetry-breaking analysis.  The percentage accuracies compare favorably with the dimensionless product of symmetry breaking factors $(\Lambda/ m_Q)$, $1/N_c$ and $\epsilon$ in each case.  (Note that the percentage accuracies of the mass combinations in Table VIII are calculated relative to the order $N_c \Lambda$ contribution to the baryon mass, i.e. $m_Q$ given in the first line of Table VII is subtracted from the mass combination, to obtain a percentage accuracy which can be compared to a product of the dimensionless factors $1/N_c$, $\left( \Lambda/m_Q \right)$ and $\epsilon$.)

In summary, there is substantive evidence for the $1/N_c$ hierarchy in the mass spectrum of heavy quark baryons.  The $1/N_c$ expansion analysis was successfull in predicting the five heavy quark baryon masses, as exhibited in Table~I.  The predictions for the remaining undiscovered bottom baryons are updated in Eq.~(\ref{predictions}).

 \vfil\break\eject

\begin{table}
\caption{\label{tab:onea} Mass predictions of Ref.~\cite{j2} and the experimentally measured masses.}
\begin{ruledtabular}
\begin{tabular*}{\textheight}{cc}
Theory & Experiment \\
\hline
$\Xi_c^\prime = 2580.8 \pm 2.1\ {\rm MeV}$ & $\Xi_c^\prime = 2576.5 \pm 2.3\ {\rm MeV}$~\cite{CLEO} \\
$\Omega_c^* = 2760.5 \pm 4.9\ {\rm MeV}$ & $\Omega_c^* = 2768.3 \pm 3.0\ {\rm MeV}$~\cite{omegacstar, pdg} \\
$\Xi_b = 5805.7 \pm 8.1\ {\rm MeV}$ & $\Xi_b^- = 5774 \pm 11 \pm 15\ {\rm MeV}$~\cite{D0xib}, 
   $\Xi_b^- = 5792.9 \pm 2.5 \pm 1.7\ {\rm MeV}$~\cite{CDFxib} \\
$\Sigma_b = 5824.2 \pm 9.0\ {\rm MeV}$ & $\Sigma_b = 5811.5 \pm 1.7\ {\rm MeV}$~\cite{CDFsigmab} \\
$\Sigma_b^* = 5840.0 \pm 8.8 \ {\rm MeV}$ & $\Sigma_b^* = 5832.7 \pm 1.8\ {\rm MeV}$~\cite{CDFsigmab}
\end{tabular*}
\end{ruledtabular}
\end{table}

\begin{table}
\caption{\label{tab:two} Mass splittings of baryons containing a single heavy quark $Q$ for
strangeness $S=0$, $-1$ and $-2$ baryons.  The operator matrix element and the $1/m_Q$ and $1/N_c$ suppressions of
each mass combination are given.  
The singlet mass combination has a mass which is $m_Q + N_c \Lambda$ at leading
order in the $1/m_Q$ and $1/N_c$ expansions.  The $J_\ell \cdot J_Q$ operator 
violates heavy quark spin symmetry and is proportional to $1/m_Q$.  
The final two columns list the 
experimental values of the mass combinations in MeV for $Q=c$ and $Q=b$ baryons
when known.}
\begin{ruledtabular}
\begin{tabular*}{\textheight}{ccccccc}
Operator & Mass Combination & $\langle O \rangle$ & {$1/m_Q$} & {$1/N_c$} & {$Q=c$} & {$Q=b$}
\\
\hline
$\openone$ & $\Lambda_Q$ & 1 & * & * & $2286.46 \pm 0.14$  & $5620.2 \pm 1.6$ \\
$J_\ell^2$ & $\frac 1 3 \left( \Sigma_Q + 2 \Sigma_Q^* \right)-\Lambda_Q$  
& 2 & $1$ & $1/N_c$ &$210.0 \pm 0.5$ & $205.4 \pm 2.1$ \\
$J_\ell \cdot J_Q$ & $\Sigma^*_Q - \Sigma_Q$  
& ${ 3 \over 2}$ & $2/m_Q$ & $1/N_c$ & $64.4 \pm 0.8$ & $21.2 \pm 2.5$ \\
\hline
$\openone$ & $\Xi_Q$ & 1 & * & * & $2469.5 \pm 0.3$ & $5792.9 \pm 3.0$ \\
$J_\ell^2$ & 
$\frac 1 3 \left( \Xi_Q^\prime + 2 \Xi_Q^* \right)-\Xi_Q$ &2   
& $1$ & $1/N_c$ & $153.7 \pm 0.9$ & \\
$J_\ell \cdot J_Q$ & $\Xi^*_Q - \Xi^\prime_Q$ & ${3 \over 2}$  
& $2/m_Q$ & $1/N_c$ & $69.5 \pm 2.3$ & \\
\hline
%
$\openone + J_\ell^2$ & $\frac 1 3 \left( \Omega_Q + 2 \Omega_Q^* \right)$ & * & * & * & $2744.7 \pm 2.2$ & \\
$J_\ell \cdot J_Q$ & $\Omega^*_Q - \Omega_Q $ & ${3 \over 2}$  
& $2/m_Q$ & $1/N_c$ & $70.8 \pm 1.5$ & \\
\end{tabular*}
\end{ruledtabular}
\end{table}

\begin{table}
\caption{\label{tab:three} Mass splittings between baryons containing a single bottom quark 
and baryons containing a single charm quark for strangeness 
$S=0$ and $S=-1$ baryons. 
The experimental values in MeV of the mass combinations 
are given in the final column.}
\begin{ruledtabular}
\begin{tabular*}{\textheight}{cccccc}
Operator & Mass Combination & $\langle O \rangle$ & {$1/m_Q$} & {$1/N_c$} & Expt. 
\\
\hline
$I_h^3$ & $\left(\Lambda_b - \Lambda_c\right)$ & 1  
& $(m_b - m_c )$ & 1 & $3333.7 \pm 1.6$ \\
$I_h^3 J_\ell^2$ & $\left[ \frac 1 3 \left( \Sigma_c+ 2 \Sigma_c^*
\right) - \Lambda_c \right] - \left[ \frac 1 3 \left( \Sigma_b+ 2 \Sigma_b^*
\right) - \Lambda_b \right]$ & 2  
& $\left( {1 \over {2m_c}} - {1 \over {2m_b}}\right)$ & $1/N_c^2$ & $4.6 \pm 2.2$  \\
\hline
$I_h^3$ & $\left( \Xi_b - \Xi_c \right)$ & 1  
& $(m_b - m_c )$ & 1 & $3323.4 \pm 3.0$  \\ 
$I_h^3 J_\ell^2$ & $\left[ \frac 1 3 \left( \Xi^\prime_c 
+ 2 \Xi_c^* \right) - \Xi_c \right] 
-\left[ \frac 1 3 \left( \Xi^\prime_b
+ 2 \Xi_b^* \right) - \Xi_b \right] $ & 2  
& $\left( {1 \over {2m_c}} - {1 \over {2m_b}}\right)$ & $1/N_c^2$ &  \\
\end{tabular*}
\end{ruledtabular}
\end{table}

\begin{table}
\caption{\label{tab:four} Mass splittings between baryons containing a single heavy quark $Q$
and baryons containing no heavy quarks for strangeness 
$S=0$ and $S=-1$ baryons. 
The experimental values in MeV of the mass combinations for $Q=c$ and $Q=b$ baryons
are given in the final two columns.}
\begin{ruledtabular}
\begin{tabular*}{\textheight}{ccccccc}
Operator & Mass Combination & $\langle O \rangle$ & {$1/m_Q$} & {$1/N_c$} & {$Q=c$} &  {$Q=b$} 
\\
\hline
$N_Q$ & $\Lambda_Q - \frac 1 4 \left( 5 N - \Delta \right)$  
& 1 & $m_Q$ & 1 & $1420.9 \pm 0.2$ & $4754.6 \pm 1.6$ \\
$N_Q J_\ell^2$ & $\left[ \frac 1 3 \left( \Sigma_Q + 2 \Sigma_Q^*
\right) - \Lambda_Q \right] - \frac 2 3 \left( \Delta - N \right)$  
& 2 & $1$ & $1/N_c^2$ & $14.4 \pm 0.8 $ & $9.8 \pm 2.2$ \\
\hline
$N_Q$ & $\Xi_Q 
- \frac 1 4 \left[ \frac 5 4 \left( 3 \Sigma + \Lambda \right) 
- \Sigma^*\right]$  
& 1 & $m_Q$ & 1 & $1348.4 \pm 0.3$ & $4671.8 \pm 3.0$ \\ 
$N_Q J_\ell^2$ & $\left[ \frac 1 3 \left( \Xi^\prime_Q 
+ 2 \Xi_Q^* \right) - \Xi_Q \right] 
- \frac 2 3 \left[ \Sigma^* 
- \frac 1 4 \left( 3\Sigma + \Lambda \right)\right]$  
& 2 & $1$ & $1/N_c^2$ & $13.2 \pm 0.9$ & \\
\end{tabular*}
\end{ruledtabular}
\end{table}

\begin{turnpage}

\begin{table}
\caption{\label{tab:five} Mass splittings of baryons containing a single heavy quark $Q$.
The operator matrix element and the 
$1/m_Q$, $1/N_c$ and $SU(3)$ flavor-breaking $\epsilon$ suppression factors of each mass combination are tabulated.    
The second set of mass combinations violate heavy quark spin symmetry, 
whereas the first set do not.
The singlet operator is
$m_Q + N_c \Lambda$ at leading orders in the $1/m_Q$ and $1/N_c$ expansions.  
Experimental values of the charm and bottom baryon mass
combinations in MeV are given in the final two columns.}
\begin{ruledtabular}
\begin{tabular*}{\textheight}{cccccccc}
Operator 
& Mass Combination & $\langle O \rangle$ & {$1/m_Q$} & {$1/N_c$} &
$\epsilon$ & {$Q=c$} & {$Q=b$} 
\\
\hline
$\openone$ 
& $\frac 1 3 \left(\Lambda_Q + 2 \Xi_Q \right)$  
& $1$ & * & * & $1$ & $2408.5 \pm 0.2$ & $5735.3 \pm 2.1$ \\
$J_\ell^2$ 
& $-\frac 1 3 \left(\Lambda_Q + 2 \Xi_Q \right)
+ \frac 1 {6} \left[ \left( \Sigma_Q + 2 \Sigma_Q^* \right)
+ \frac 2 3 \left( \Xi^\prime_Q + 2 \Xi_Q^* \right) 
+ \frac 1 3 \left( \Omega_Q  + 2 \Omega_Q^* \right) \right]$  
& $2$ & $1$ & $1/N_c$ & $1$ & $171.6 \pm 0.5$ & \\
$T^8$ 
& $\left( \Xi_Q - \Lambda_Q \right)$  
& $\sqrt{3} \over 2$ & $1$ & $1$ & $\epsilon$ & $183.0 \pm 0.3 $ & $ 172.7 \pm 3.4$ \\
$J_\ell^i G^{i8}$ 
& $-\frac 5 8
\left( \Lambda_Q - \Xi_Q \right)+ \frac 1 {8} \left[  \left( \Sigma_Q
+ 2 \Sigma_Q^* \right)  - \frac 1 3 \left( \Xi^\prime_Q + 2 \Xi_Q^* \right)
-\frac 2 3\left( \Omega_Q + 2 \Omega_Q^* \right) \right] $    
& ${1 \over {2 \sqrt{3}}}{{15} \over 8} $ & $1$ & $1/N_c$ & $\epsilon$ & $36.5 \pm 0.6$ & \\
$\{T^8,T^8\}$ 
& $\frac 1 {2} \left[ \frac 1 3 \left( \Sigma_Q + 2 \Sigma_Q^* \right)
- \frac 2 3 \left( \Xi^\prime_Q + 2 \Xi_Q^* \right)
+ \frac 1 3 \left( \Omega_Q + 2 \Omega_Q^* \right) \right] $  
& $3 \over 2$ & $1$ & $1/N_c$ & $\epsilon^2$ & $-2.6 \pm 1.4$ & \\
\hline
$J_\ell \cdot J_Q$ 
& $\frac 1 6 \left[ 3 \left( \Sigma^*_Q - \Sigma_Q \right)
+ 2 \left( \Xi^*_Q - \Xi_Q^\prime \right) 
+ \left( \Omega^*_Q - \Omega_Q \right) \right]$   
& $ 3 \over 2$ & $2/m_Q$ & $1/N_c$ & $1$ & $67.2 \pm 0.9$ & \\
$J_Q^i G^{i8}$ 
& $\Xi_Q \Xi^\prime_Q$  
& $3 \over 8$ & $2/m_Q$ & $1/N_c$ & $\epsilon$ & $ $ & \\
$\left( J_\ell \cdot J_Q \right) T^8$ 
& $-\frac 1 {6} \left[ 3 \left( \Sigma^*_Q - \Sigma_Q \right) 
- \left( \Xi^*_Q - \Xi^\prime_Q \right)
-2 \left( \Omega^*_Q - \Omega_Q \right) \right] 
+\frac 5 {2\sqrt{3}} \Xi_Q \Xi^\prime_Q $   
& ${1 \over {2 \sqrt{3}}} {{15} \over 4}$ & $2/m_Q$ & $1/N_c^2$ & $\epsilon$ &  & \\
$J_Q^i \{T^8,G^{i8}\}$ 
& $\frac 1 4 \left[ \left( \Sigma^*_Q - \Sigma_Q \right)
- 2 \left( \Xi^*_Q - \Xi_Q^\prime \right) 
+ \left( \Omega^*_Q - \Omega_Q \right) \right]$   
& $ 9 \over {16}$ & $2/m_Q$ & $1/N_c^2$ & $\epsilon^2$ & $-0.9 \pm 1.2$ & \\
\hline
%
 & 
$-\frac 1 {6} \left[ 3 \left( \Sigma^*_Q - \Sigma_Q \right) 
- \left( \Xi^*_Q - \Xi^\prime_Q \right)
-2 \left( \Omega^*_Q - \Omega_Q \right) \right]$ 
& * & $2/m_Q$ & $1/N_c$ & $\epsilon$ & $2.6 \pm 0.8$ & \\
\end{tabular*}
\end{ruledtabular}
\end{table}
\end{turnpage}

\begin{turnpage}

\begin{table}
\caption{\label{tab:six} Mass splittings which violate heavy quark flavor symmetry, but preserve heavy quark spin symmetry.  Experimental values of the mass
combinations in MeV are given in the final column.}
\begin{ruledtabular}
\begin{tabular*}{\textheight}{cccccccc}
Operator 
& Mass Combination & $\langle O \rangle$ & {$1/m_Q$} & {$1/N_c$} &
$\epsilon$ & Expt.
\\
\hline
$I_h^3$ 
& $\frac 1 3 \left(\Lambda_b+ 2 \Xi_b \right)-\frac 1 3 \left(\Lambda_c+ 2 \Xi_c \right)$  
& 1 & $(m_b - m_c )$ & 1 & $1$ & $3326.8 \pm 2.1$  \\
$I_h^3 J_\ell^2$ 
& $\left\{-\frac 1 3 \left(\Lambda_c 
+ 2 \Xi_c \right)+ \frac 1 {6} \left[  \left(\Sigma_c + 2 \Sigma_c^*\right) 
+ \frac 2 3 \left( \Xi^\prime_c + 2 \Xi^*_c \right) 
+ \frac 1 3 \left( \Omega_c + 2 \Omega_c^*\right)\right]\right\}$  
& &  &  &  &   \\
%
 & $-\left\{-\frac 1 3 \left(\Lambda_b 
+ 2 \Xi_b \right)+ \frac 1 {6} \left[  \left(\Sigma_b + 2 \Sigma_b^*\right) 
+ \frac 2 3 \left( \Xi^\prime_b + 2 \Xi^*_b \right) 
+ \frac 1 3 \left( \Omega_b + 2 \Omega_b^*\right)\right]\right\}$  
& 2 & $\left( { 1 \over {2m_c}} - {1 \over {2m_b}} \right)$ & $1/N_c^2$ & $1$ &   \\
$I_h^3 T^8$ 
& $\left(\Xi_c -  \Lambda_c  \right) - \left( \Xi_b - \Lambda_b  \right)$  
& $\sqrt{3} \over 2$ & $\left( { 1 \over {2m_c}} - {1 \over {2m_b}} \right)$ & $1/N_c$ & $\epsilon$ & $10.3 \pm 3.4$  \\
$I_h^3 J_\ell^i G^{i8}$ 
& $\left\{\frac 5 8
\left( \Lambda_c - \Xi_c \right)- \frac 1 {8} \left[ \left(\Sigma_c + 2
\Sigma_c^*\right) -\frac 1 3  \left(\Xi^\prime_c + 2 \Xi_c^* \right)
-\frac 2 3 \left(\Omega_c + 2 \Omega_c^*\right)\right]\right\}$  
&  &  &  &  &  \\
%
& $-\left\{\frac 5 8
\left( \Lambda_b - \Xi_b \right)- \frac 1 {8} \left[ \left(\Sigma_b + 2
\Sigma_b^*\right) -\frac 1 3  \left(\Xi^\prime_b + 2 \Xi_b^* \right)
-\frac 2 3 \left(\Omega_b + 2 \Omega_b^*\right)\right]\right\}$ 
& ${1 \over {2\sqrt{3}}} {{15} \over 8}$ & $\left( { 1 \over {2m_c}} - {1 \over {2m_b}} \right)$ & $1/N_c^2$ & $\epsilon$ &  \\
$I_h^3 \{T^8,T^8\}$ 
& $\frac 1 2 \left[\frac 1 3
\left( \Sigma_c + 2
\Sigma_c^* \right)- \frac 2 3 \left(\Xi^\prime_c + 2 \Xi_c^* \right) +
\frac 1 3 \left(\Omega_c + 2 \Omega_c^* \right) \right]$  
& & & & & \\
%
&  $-\frac 1 2 \left[\frac 1 3 \left( \Sigma_b + 2
\Sigma_b^* \right)- \frac 2 3 \left(\Xi^\prime_b + 2 \Xi_b^* \right) +
\frac 1 3 \left(\Omega_b + 2 \Omega_b^* \right) \right]$
& $ 3 \over 2$ & $\left( { 1 \over {2m_c}} - {1 \over {2m_b}} \right)$ & $1/N_c^2$ & $\epsilon^2$ &  \\
\end{tabular*}
\end{ruledtabular}
\end{table}
\end{turnpage}

\begin{turnpage}
\begin{table}
\caption{\label{tab:seven} Mass splittings between baryons
containing a single heavy quark $Q$ and baryons containing no heavy quark.  Each
$N_Q$-dependent operator corresponds to a difference of a $Qqq$ baryon mass splitting
and a $qqq$ baryon mass splitting.  Experimental values of the mass
combinations in MeV for $Q=c$ and $Q=b$ are given in the final two columns.}
\begin{ruledtabular}
\begin{tabular*}{\textheight}{cccccccc}
Operator 
& Mass Combination & $ \langle O \rangle$ & {$1/m_Q$} & {$1/N_c$} &
$\epsilon$ & {$Q=c$} & $Q=b$
\\
\hline
$N_Q$ 
& $\frac 1 3 \left(\Lambda_Q + 2 \Xi_Q \right)
- \frac 1 4 \left[ \frac 5 8 \left( 2 N + 3 \Sigma + \Lambda + 2 \Xi \right) 
-\frac 1 {10} \left( 4 \Delta + 3 \Sigma^* + 2 \Xi^* +\Omega \right) \right]$  
& $1$ & $m_Q$ & 1 & $1$ & $1315.1 \pm 0.2$ & $4641.9 \pm 2.1$ \\
$N_Q J_\ell^2$ 
& $\left\{-\frac 1 3 \left(\Lambda_Q 
+ 2 \Xi_Q \right)+ \frac 1 {6} \left[  \left(\Sigma_Q + 2 \Sigma_Q^*\right) 
+ \frac 2 3 \left( \Xi^\prime_Q + 2 \Xi^*_Q \right) 
+ \frac 1 3 \left( \Omega_Q + 2 \Omega_Q^*\right)\right]\right\}$  
&  &  &  &  &  & \\
%
  & $- \frac 2 3 \left[ \frac 1 {10} \left( 4 \Delta + 3 \Sigma^* + 2
\Xi^* + \Omega \right) - \frac 1 8 \left( 2 N + 3 \Sigma + \Lambda + 2 \Xi
\right) \right]$  
& 2 & $1$ & $1/N_c^2$ & $1$ & $17.6 \pm 0.6 $ & \\
$N_Q T^8$ 
& $-\left( \Xi_Q - \Lambda_Q \right)- \frac 1 {8}
\left( 6 N - 3 \Sigma +  \Lambda - 4 \Xi \right) + \frac 1 {20} \left( 2 \Delta
- \Xi^* - \Omega \right)$  
& $\sqrt{3} \over 2$ & $1$ & $1/N_c$ & $\epsilon$ & $42.9 \pm 0.3 $ & $53.2 \pm 3.4$ \\
$N_Q J_\ell^i G^{i8}$ 
& $\left\{\frac 5 8
\left( \Lambda_Q - \Xi_Q \right)- \frac 1 {8} \left[ \left(\Sigma_Q + 2
\Sigma_Q^*\right) -\frac 1 3  \left(\Xi^\prime_Q + 2 \Xi_Q^* \right)
-\frac 2 3 \left(\Omega_Q + 2 \Omega_Q^*\right)\right]\right\}$  
& &  &  &  &  & \\
%
  & $-\frac 1 {24} \left( 8 N - 9 \Sigma + 3 \Lambda -2 \Xi \right)
+ \frac 1 {12} \left( 2 \Delta - \Xi^* - \Omega \right)$  
& ${1 \over {2 \sqrt{3}}} {{15} \over 8}$ & $1$ & $1/N_c^2$ & $\epsilon$ & $6.6 \pm 0.6$ & \\
$N_Q \{T^8,T^8\}$ 
& $\frac 1 2 \left[\frac 1 3
\left( \Sigma_Q + 2
\Sigma_Q^* \right)- \frac 2 3 \left(\Xi^\prime_Q + 2 \Xi_Q^* \right) +
\frac 1 3 \left(\Omega_Q + 2 \Omega_Q^* \right) \right]$  
& &  &  &  &  & \\
%
  & $+\frac 1 3 \left[ -\frac 1 4 \left( 2 N - \Sigma -3 \Lambda + 2 \Xi
\right) - \frac 1 7 \left( 4 \Delta - 5 \Sigma ^* -2 \Xi^* + 3 \Omega \right) \right]$  
& $3 \over 2$ & $1$ & $1/N_c^2$ & $\epsilon^2$ & $ 1.6 \pm 1.4$ & \\
\end{tabular*}
\end{ruledtabular}
\end{table}
\end{turnpage}

\begin{turnpage}
\begin{table}
\caption{\label{tab:eight} Accuracy of $Q=c$ and $Q=b$ baryon mass splittings 
relative to the order $N_c$ mass of each mass combination.  
}
\begin{ruledtabular}
\begin{tabular*}{\textheight}{cccccc}
Mass Combination  & {$1/m_Q$} & {$1/N_c$} & Flavor & {$Q=c$} & {$Q=b$} \\
\hline
$-\frac 1 3 \left(\Lambda_Q + 2 \Xi_Q \right)
+ \frac 1 {6} \left[ \left( \Sigma_Q + 2 \Sigma_Q^* \right)
+ \frac 2 3 \left( \Xi^\prime_Q + 2 \Xi_Q^* \right) 
+ \frac 1 3 \left( \Omega_Q  + 2 \Omega_Q^* \right) \right]$  
& $1$ & $1/N_c^2$ & $1$ & $14.55 \pm 0.04 \%$ & \\
$\left( \Xi_Q - \Lambda_Q \right)$  
& $1$ & $1/N_c$ & $\epsilon$ & $17.22 \pm 0.03 \%$ & $16.22 \pm 0.32 \%$ \\
$-\frac 5 8
\left( \Lambda_Q - \Xi_Q \right)+ \frac 1 {8} \left[  \left( \Sigma_Q
+ 2 \Sigma_Q^* \right)  - \frac 1 3 \left( \Xi^\prime_Q + 2 \Xi_Q^* \right)
-\frac 2 3\left( \Omega_Q + 2 \Omega_Q^* \right) \right] $    
& $1$ & $1/N_c^2$ & $\epsilon$ & $3.18 \pm 0.05 \%$ & \\
$-\frac 1 {2} \left[ \frac 1 3 \left( \Sigma_Q + 2 \Sigma_Q^* \right)
- \frac 2 3 \left( \Xi^\prime_Q + 2 \Xi_Q^* \right)
+ \frac 1 3 \left( \Omega_Q + 2 \Omega_Q^* \right) \right] $  
& $1$ & $1/N_c^2$ & $\epsilon^2$ & $0.20 \pm 0.11 \%$ & \\
\hline
$\frac 1 6 \left[ 3 \left( \Sigma^*_Q - \Sigma_Q \right)
+ 2 \left( \Xi^*_Q - \Xi_Q^\prime \right) 
+ \left( \Omega^*_Q - \Omega_Q \right) \right]$   
& $1/m_Q$ & $1/N_c^2$ & $1$ & $5.36 \pm 0.07 \%$ & \\
$\Xi_Q \Xi^\prime_Q$  
& $1/m_Q$ & $1/N_c^2$ & $\epsilon$ & $ $ & \\
$-\frac 1 {6} \left[ 3 \left( \Sigma^*_Q - \Sigma_Q \right) 
- \left( \Xi^*_Q - \Xi^\prime_Q \right)
-2 \left( \Omega^*_Q - \Omega_Q \right) \right]
+\frac 5 {2\sqrt{3}} \Xi_Q \Xi^\prime_Q $   
& $1/m_Q$ & $1/N_c^3$ & $\epsilon$ & & \\
$\frac 1 4 \left[ \left( \Sigma^*_Q - \Sigma_Q \right)
- 2 \left( \Xi^*_Q - \Xi_Q^\prime \right) 
+ \left( \Omega^*_Q - \Omega_Q \right) \right]$   
& $1/m_Q$ & $1/N_c^3$ & $\epsilon^2$ & $-0.07 \pm 0.05 \%$ & \\
\hline
$-\frac 1 {6} \left[ 3 \left( \Sigma^*_Q - \Sigma_Q \right) 
- \left( \Xi^*_Q - \Xi^\prime_Q \right)
-2 \left( \Omega^*_Q - \Omega_Q \right) \right]$   
& $1/m_Q$ & $1/N_c^2$ & $\epsilon$ & $0.20 \pm 0.06 \%$ & \\
\end{tabular*}
\end{ruledtabular}
\end{table}
\end{turnpage}

\end{document}